\documentclass[aps,PRB,twocolumn,superscriptaddress,preprintnumbers]{revtex4-2}
\usepackage[T1]{fontenc}
\usepackage{times}
\usepackage{graphicx,color}
\usepackage{amsfonts,amsmath,amssymb,amsbsy}
\usepackage[colorlinks=true, linkcolor=blue, citecolor=blue, urlcolor=blue]{hyperref}
\usepackage{float}

\begin{document}

\title{Spin-selective elliptic optical dichroism and \\
perfectly spin-polarized third-order nonlinear photocurrent in altermagnets}
\author{Motohiko Ezawa}
\affiliation{Department of Applied Physics, The University of Tokyo, 7-3-1 Hongo, Tokyo
113-8656, Japan}

\begin{abstract}
We show that the low-energy theory of a $d$-wave altermagnet is
characterized by anisotropic Dirac cones with spin-split band structures
based on a recently proposed tight-binding model. In this system,
spin-selective perfect elliptic dichroism emerges, enabling exclusive
excitation of either up-spin or down-spin electrons by tuning the
ellipticity of the incident light. We further derive a formula for the
third-order photocurrent generated under simultaneous application of
elliptically polarized light and a static electric field, expressed in terms
of the quantum metric and Berry curvature. Using this formula, we
demonstrate that only an up-spin polarized current is induced. This
third-order response constitutes the leading photocurrent, as second-order
processes such as injection and shift currents are forbidden by the
inversion symmetry inherent to altermagnets.  They lead to the prediction
that, in insulating $d$-wave altermagnets, a spin-polarized third-order
photocurrent arises as the leading electric response when elliptically
polarized light and a static electric field are applied simultaneously.
These findings provide a foundation for future developments in photoinduced
spintronics based on altermagnets.
\end{abstract}

\date{\today }
\maketitle

\textbf{Introduction: }Altermagnets\cite{SmejX,SmejX2} are one of the
hottest topics in condensed matter physics because they will be useful for
future spintronics applications. They are collinear antiferromagnets but
break time reversal symmetry, which allows nonzero anomalous Hall effect.
Especially, $d$-wave altermagnets are prominent, where spin currents are
induced by electric field\cite{Naka,Gonza,NakaB,Bose,NakaRev,APEX} and
temperature gradient\cite{Naka,APEX}. $d$-wave altermagnets have a
combinatorial symmetry $\mathcal{C}_{4z}\mathcal{T}$  made of
time-reversal symmetry and the four-fold rotational symmetry\cite%
{SmejX,SmejX2}. This symmetry enforces a spin-split band structure,  %
where spins are opposite by the 90-degree momentum rotation.  In
addition, altermagnets have inversion symmetry. There is a simple four-band
tight-binding model possessing all these symmetry properties of $d$-wave
altermagnets and confirmed by the density-functional theory\cite{ZFu}.

Circular dichroism is an intriguing phenomenon, which was originally
proposed in honeycomb systems\cite{Yao08,Xiao,EzawaOpt,Li}. The
tight-binding model based on the honeycomb lattice possesses two Dirac cones
assigned by different valley degrees of freedom. By irradiating
circularly-polarized light to the system, only electrons at one valley are
excited. Furthermore, circular dichroism is generalized to elliptic
dichroism for anisotropic Dirac cones\cite{TCIOpt,EllipticP,APEX}, where
only one valley is excited by irradiating elliptically-polarized light.
Elliptic dichroism is formulated in terms of quantum metric and the Berry
curvature\cite{EllipticP,APEX}.

The leading photoinduced current is the second order in electric field,
i.e., $j\propto E^{2}$, where there are injection current\cite%
{Sipe,JuanNC,Juan,Ave,AhnX,AhnNP,WatanabeInject,Okumura,Dai,EzawaVolta} and
the shift current\cite%
{Young,Young2,Ave,Kraut,Baltz,Sipe,Juan,AhnX,MorimotoScAd,Kim,Barik,AhnNP,WatanabeInject,Dai,Yoshida,EzawaVolta}%
. However, they are prohibited if the system has inversion symmetry because
current and electric field are odd under inversion symmetry. Hence, there is
no second-order photocurrent in altermagnets. On the other hand, the
third-order photocurrent $j\propto E^{3}$\ is allowed even in the presence
of inversion symmetry. A typical example is the jerk current\cite%
{Jerk,JerkComment,JerkReply,snap,PJerk}, which has been discussed so far
only for linearly polarized light.

In this paper, we derive a low-energy theory of altermagnets characterized
by anisotropic Dirac cones.  Based on this theory, we demonstrate that
elliptic dichroism arises in altermagnets, whereby electrons with either up
or down spin can be selectively excited by appropriately tuning the
ellipticity of the incident light. Furthermore, we derive, for the first
time, a formula for higher-order photocurrents under elliptically polarized
light, expressed in terms of the quantum metric and Berry curvature,
analogous to the formula for elliptic optical dichroism. We then show that a
perfectly spin-polarized third-order photocurrent is generated when
elliptically polarized light and a static electric field are applied
simultaneously. If the system is an insulator, the third-order photocurrent
is the leading-order current. Our findings provide a foundation for future
developments in photo-excited spintronics based on altermagnets.

\begin{figure}[t]
\centerline{\includegraphics[width=0.49\textwidth]{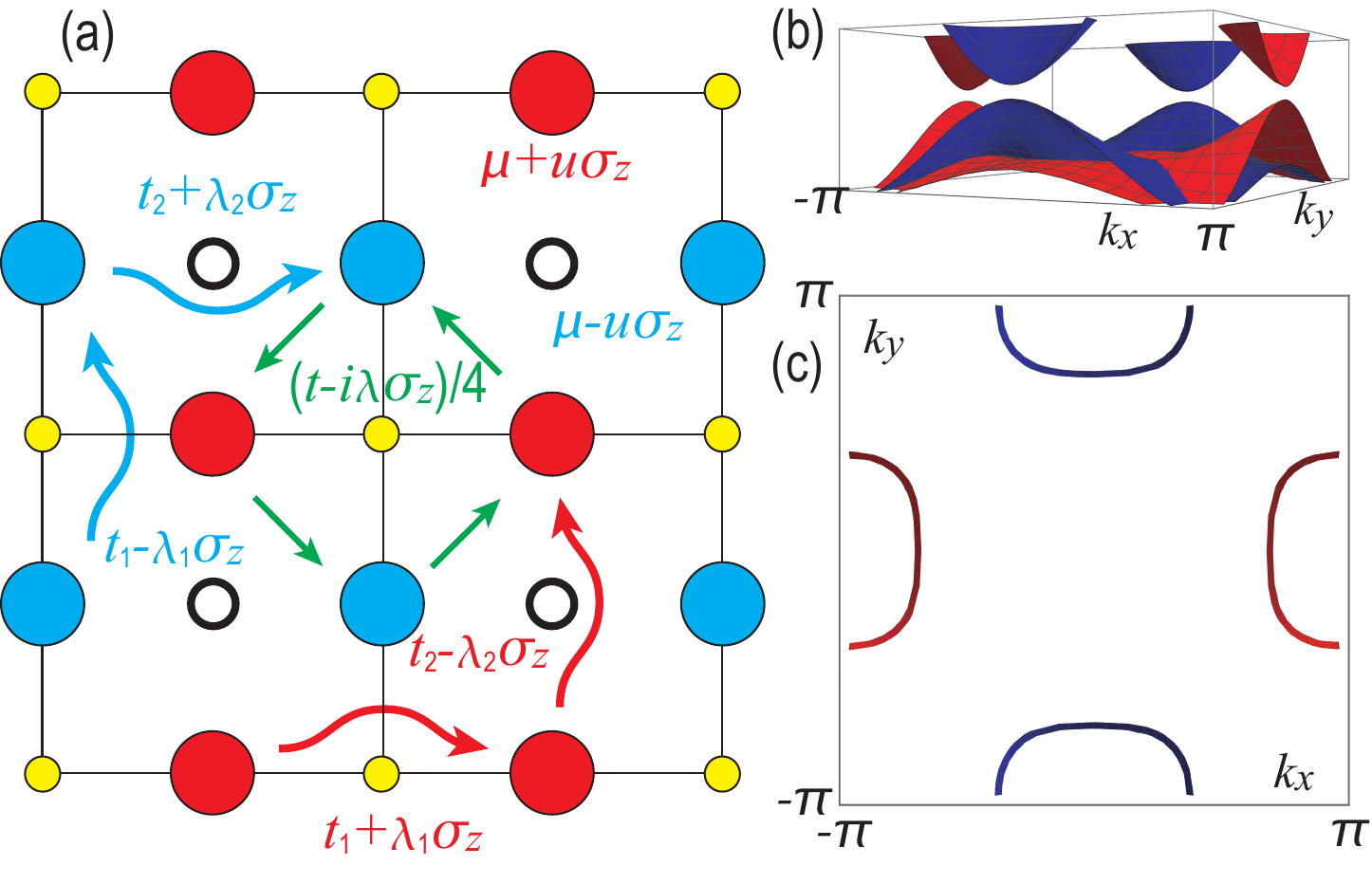}}
\caption{(a) Illustration of the Lieb lattice model. Red (cyan) disks
represent sites with up (down) spin. Small yellow disks represent
non-magnetic sites. The tight-binding model is constructed only with the use
of red and blue disks, where the hopping terms are modified in the presence
of yellow disks. Red (cyan) arrows represent hoppings between the
next-nearest neighbor up(down)-spin sites. Green arrows represent hoppings
between the nearest-neighbor sites between opposite spins. We have set $%
A=t_{1}+t_{2}$, $B=t_{1}-t_{2}$,\ $C=\protect\lambda _{1}+\protect\lambda %
_{2}$\ and $D=\protect\lambda _{1}-\protect\lambda _{2}$ in the Hamiltonian (%
\protect\ref{H1}). (b) Bird's eye's view of the band structure containing
two anisotropic Dirac cones. (c) Top view of the band structure. Red (blue)
color represents the up-spin (down-spin) band structure.}
\label{FigLieb}
\end{figure}

\textbf{Model:} We start with the four-band tight-bind model for $d$-wave
altermagnets defined on the square lattice\cite{ZFu,Anton,PFeng},%
\begin{align}
H\left( k_{x},k_{y}\right) =& A\left( \cos ak_{x}+\cos ak_{y}\right) \tau
_{0}\sigma _{0}  \notag \\
& +B\left( \cos ak_{x}-\cos ak_{y}\right) \tau _{z}\sigma _{0}  \notag \\
& +t\cos \frac{ak_{x}}{2}\cos \frac{ak_{y}}{2}\tau _{x}\sigma _{0}+\lambda
\sin \frac{ak_{x}}{2}\sin \frac{ak_{y}}{2}\tau _{y}\sigma _{z}  \notag \\
& +C\left( \cos ak_{x}-\cos ak_{y}\right) \tau _{0}\sigma _{z}  \notag \\
& +\left( u+D\left( \cos ak_{x}+\cos ak_{y}\right) \right) \tau _{z}\sigma
_{z},  \label{H1}
\end{align}%
 where $\sigma _{j}$ is the Pauli matrix for the spin, and $\tau _{j}$ is
that for the sublattice degrees of freedom with $j=x,y,z$. $a$ is the
lattice constant, and $u$ is a real number proportional to the spin
order. $A$ and $B$ are the isotropic and anisotropic hoppings of the
next-nearest neighbor sites, while $C$ and $D$ are the isotropic and
anisotropic spin-dependent hoppings of the next-nearest neighbor sites. $t$
is the hopping and $\lambda $ is the spin-dependent hopping of
nearest-neighbor sites. We assume $t>0$ and $\lambda >0$. The model was
derived from the Lieb lattice as shown in Fig.\ref{FigLieb}(a).  The
Hamiltonian has such a symmetry\cite{ZFu} that $\mathcal{C}_{4z}\mathcal{T}=%
\frac{i}{\sqrt{2}}\left( \sigma _{y}-\sigma _{x}\right) \tau _{x}\mathcal{K}$
with%
\begin{equation}
\mathcal{C}_{4z}\mathcal{T}H\left( k_{x},k_{y}\right) \mathcal{C}_{4z}%
\mathcal{T}=H\left( -k_{y},k_{x}\right) ,
\end{equation}%
which guarantees the existence of opposite spin-split band structure by
90-degree rotation of the momentum, $\mathcal{C}_{4z}\mathcal{T}\left(
k_{x},k_{y}\right) \mathcal{C}_{4z}\mathcal{T}=\left( -k_{y},k_{x}\right) $.%

Because $\sigma _{z}$ is a good quantum number, we may set $\sigma _{z}=s$
with $s=\pm 1$ in the Hamiltonian (\ref{H1}). Then, the Hamiltonian is
reduced to a direct sum of two two-band models, $H=H_{\uparrow }\oplus
H_{\downarrow }$. The band structure is shown in Fig.\ref{FigLieb}(b) and
(c). With respect to the band with up spin $s=1$, there is a Dirac cone at
the X point, where the Hamiltonian is expanded as%
\begin{equation}
H_{\uparrow }^{\text{X}}=-4C+tak_{x}^{\prime }\tau _{x}+\lambda ak_{y}\tau
_{y}-\left( 4B-2u\right) \tau _{z}
\end{equation}%
 with $k_{x}^{\prime }\equiv k_{x}-\pi /a$.  It represents an
anisotropic Dirac cone. Similarly, with respect to the band with down spin $%
s=-1$, there is a Dirac cone at the Y point, where the Hamiltonian is
expanded as%
\begin{equation}
H_{\downarrow }^{\text{Y}}=-4C+tak_{y}^{\prime }\tau _{x}-\lambda ak_{x}\tau
_{y}+\left( 4B-2u\right) \tau _{z}
\end{equation}%
 with $k_{y}^{\prime }\equiv k_{y}-\pi /a$. The two valleys are
connected by the $\mathcal{C}_{4z}\mathcal{T}$ symmetry, where spins are
opposite between two valleys.

The Hamiltonians $H_{\uparrow }^{\text{X}}$ and $H_{\downarrow }^{\text{Y}}$
are rewritten in the form of $H\left( \mathbf{k}\right) =h_{0}\left( \mathbf{%
k}\right) +\mathbf{h}\left( \mathbf{k}\right) \cdot \mathbf{\tau }$. The
energy spectrum is obtained as%
\begin{equation}
\varepsilon _{\pm }\left( \mathbf{k}\right) =h_{0}\left( \mathbf{k}\right)
\pm \sqrt{\left\vert \mathbf{h}\left( \mathbf{k}\right) \right\vert }.
\label{ek}
\end{equation}%
We define the normalized vector $\mathbf{n}\left( \mathbf{k}\right) $ as $%
\mathbf{n}\left( \mathbf{k}\right) \equiv \mathbf{h}\left( \mathbf{k}\right)
/|\mathbf{h}\left( \mathbf{k}\right) |=\left( \sin \theta \cos \phi ,\sin
\theta \sin \phi ,\cos \theta \right) $. The quantum metric for the two-band
system is explicitly given\cite%
{Matsuura,Bleu,Gers,Robin,OnishiX,WChen2024,EzawaQG,Oh} by $g_{\pm }^{\mu
\nu }\left( \mathbf{k}\right) =\pm \frac{1}{4}\left( \partial _{k_{\mu }}%
\mathbf{n}\right) \cdot \left( \partial _{k_{\nu }}\mathbf{n}\right) $ with
the normalized vector $\mathbf{n}\left( \mathbf{k}\right) $.

The Berry curvature for the two-band system is explicitly given\cite%
{Hsiang,Stic,Jiang,Bleu,Robin} by $\Omega _{\pm }^{xy}\left( \mathbf{k}%
\right) =\mp \frac{1}{2}\mathbf{n}\cdot \left( \partial _{k_{x}}\mathbf{n}%
\times \partial _{k_{y}}\mathbf{n}\right) $. It is a solid angle spanned by
the vector $\mathbf{n}$. The quantum metrices $g_{xx}^{\text{X}}\left( 
\mathbf{k}\right) $, $g_{yy}^{\text{X}}\left( \mathbf{k}\right) $ and the
Berry curvature $\Omega _{xy}^{\text{X}}\left( \mathbf{k}\right) $ at the X
point are given by\ 
\begin{align}
g_{xx}^{\text{X}}\left( k_{x}^{\prime },k_{y}\right) =& \frac{%
a^{2}t^{2}\left( \left( \lambda ak_{y}\right) ^{2}+\Delta ^{2}\right) }{%
4\left( \left( tak_{x}^{\prime }\right) ^{2}+\left( \lambda ak_{y}\right)
^{2}+\Delta ^{2}\right) ^{2}},  \label{g1} \\
g_{yy}^{\text{X}}\left( k_{x}^{\prime },k_{y}\right) =& \frac{a^{2}\lambda
^{2}\left( \left( tak_{x}^{\prime }\right) ^{2}+\Delta ^{2}\right) }{4\left(
\left( tak_{x}^{\prime }\right) ^{2}+\left( \lambda ak_{y}\right)
^{2}+\Delta ^{2}\right) ^{2}},  \label{g2} \\
\Omega _{xy}^{\text{X}}\left( k_{x}^{\prime },k_{y}\right) =& \frac{%
a^{2}\Delta t\lambda }{2\left( \left( tak_{x}^{\prime }\right) ^{2}+\left(
\lambda ak_{y}\right) ^{2}+\Delta ^{2}\right) ^{3/2}}  \label{omega}
\end{align}%
 with the Dirac mass $\Delta =4B-2u$. Similarly, we have%
\begin{align}
g_{xx}^{\text{Y}}\left( k_{x},k_{y}^{\prime }\right) =& g_{xx}^{\text{X}%
}\left( k_{x}^{\prime },k_{y}\right) , \\
g_{yy}^{\text{Y}}\left( k_{x},k_{y}^{\prime }\right) =& g_{yy}^{\text{X}%
}\left( k_{x}^{\prime },k_{y}\right) , \\
\Omega _{xy}^{\text{Y}}\left( k_{x},k_{y}^{\prime }\right) =& \Omega _{xy}^{%
\text{X}}\left( k_{x}^{\prime },k_{y}\right)
\end{align}%
at the Y point. The integration of the Berry curvature gives the
half-quantized Chern number for a single Dirac cone,%
\begin{equation}
\frac{1}{2\pi }\int \Omega _{xy}^{\text{X}}\left( \mathbf{k}\right) d\mathbf{%
k}=\frac{1}{2\pi }\int \Omega _{xy}^{\text{Y}}\left( \mathbf{k}\right) d%
\mathbf{k}=\frac{\Delta }{2\left\vert \Delta \right\vert }.
\end{equation}%
By summing up two Chern numbers from the two Dirac cones, we obtain a
quantized Chern number,%
\begin{equation}
\frac{1}{2\pi }\int \left( \Omega _{xy}^{\text{X}}\left( \mathbf{k}\right)
+\Omega _{xy}^{\text{Y}}\left( \mathbf{k}\right) \right) d\mathbf{k}=\frac{%
\Delta }{\left\vert \Delta \right\vert }.
\end{equation}%
It shows that the system is a Chern insulator, where quantum anomalous Hall
effects occur.

\begin{figure}[t]
\centerline{\includegraphics[width=0.49\textwidth]{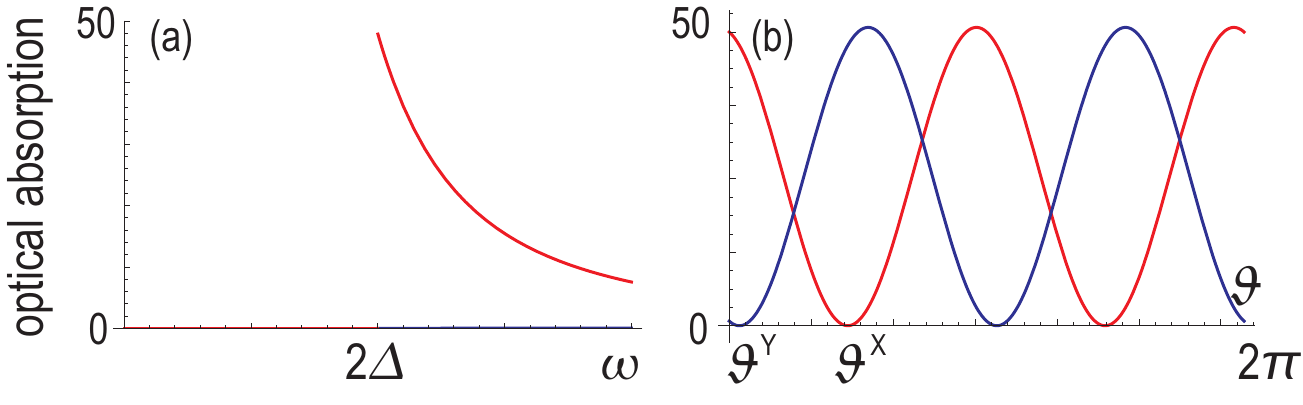}}
\caption{ (a) Optical absorption as a function of $\protect\omega $  in
units of $\hbar \protect\sigma _{0}\left\vert \Delta \right\vert /\left(
a^{2}t^{2}\right) $.  Red (blue) curve corresponds to the spin up (down)
polarization. We have set $\protect\vartheta =\protect\vartheta ^{\text{Y}}$%
, where only electrons with up spin is excited. (b) Optical absorption as a
function of $\protect\vartheta $ at the optical band edge $\hbar \protect%
\omega =2\left\vert \Delta \right\vert $. We have set $t=4;\protect\lambda %
=0.5;B=-1$ and $u=-2.2$.}
\label{FigOpt}
\end{figure}

\textbf{Elliptic optical dichroism:} We next analyze optical absorption. We
apply elliptically polarized light with the electric field%
\begin{equation}
\mathbf{E}\left( \omega \right) =E\left( \omega \right) \left( \mathbf{e}%
_{x}\cos \vartheta +i\mathbf{e}_{y}\sin \vartheta \right) ,
\end{equation}%
where $\vartheta $ is the ellipticity of injected light $0<\vartheta <\pi $
for the right polarization, and $-\pi <\vartheta <0$ for the left
polarization. $\vartheta =\pi /4$ corresponds to the right circularly
polarized right, while $\vartheta =-\pi /4$ corresponds to the left
circularly polarized right. The optical conductivity is calculated as\cite%
{EllipticP,APEX}%
\begin{equation}
\sigma \left( \omega ;\vartheta \right) =\hbar \sigma _{\text{H}}\int d%
\mathbf{k}f(\mathbf{k})g(\mathbf{k};\vartheta )\delta \left[ \varepsilon
_{+}(\mathbf{k})-\varepsilon _{-}(\mathbf{k})-\hbar \omega \right] 
\label{OptG}
\end{equation}%
with the Hall conductivity unit $\sigma _{\text{H}}\equiv e^{2}/\hbar $, and%
\begin{equation}
g(\mathbf{k};\vartheta )\equiv g_{xx}(\mathbf{k})\cos ^{2}\vartheta +g_{yy}(%
\mathbf{k})\sin ^{2}\vartheta +\Omega _{xy}(\mathbf{k})\sin \vartheta \cos
\vartheta ,  \label{gk}
\end{equation}%
where $g_{xx}(\mathbf{k})$ and $g_{yy}(\mathbf{k})$ are quantum metrices, $%
\Omega _{xy}(\mathbf{k})$ is the Berry curvature, while $f(\mathbf{r},%
\mathbf{k})=1/\left( \exp \left( \left( \varepsilon \left( \mathbf{k}\right)
-\mu \right) /\left( k_{\text{B}}T\left( \mathbf{r}\right) \right) \right)
+1\right) $ is the Fermi distribution function.  The detailed derivation
is shown in Supplemental Material I. We denote Eq.(\ref{gk}) by $g^{\text{X}}(%
\mathbf{k};\vartheta )$ at the X point and $g^{\text{Y}}(\mathbf{k}%
;\vartheta )$ at the Y point. It is observed from Eq.(\ref{OptG}) that the
minimum of the excitation frequency $\omega $ is given by the optical band
edge $\hbar \omega =2\left\vert \Delta \right\vert $, where excitation
occurs at the X or Y point: See Fig.\ref{FigOpt}(a). There is no optical
absorption below the optical band edge.  In the following, we consider
the zero-temperature limit and we take the chemical potential within the
Dirac gap, $\mu =-4C$, where the system is an insulator.

We calculate Eq.(\ref{OptG}) based on the Dirac model around the X and Y
points. Namely, we set $\mathbf{k}=\left( k_{x}^{\prime },k_{y}\right) $ at
the X point and $\mathbf{k}=\left( k_{x},k_{y}^{\prime }\right) $ at the Y
point. After integrating the momentum integration, we obtain %
\begin{equation}
\sigma \left( \omega ;\vartheta \right) =\hbar \sigma _{\text{H}%
}\int_{0}^{2\pi }\frac{k_{\text{F}}g\left( k_{\text{F}}\right) }{\left\vert
\left. \partial _{k}\varepsilon \left( k\right) \right\vert _{k_{\text{F}%
}}\right\vert }d\phi =\frac{\hbar ^{2}\omega \sigma _{\text{H}}}{2a^{2}t^{2}}%
\int_{0}^{2\pi }g\left( k_{\text{F}}\right) d\phi ,  \label{OptC}
\end{equation}%
where $ak_{\text{F}}=\sqrt{\left( \hbar \omega /2\right) ^{2}-\Delta ^{2}}/t$
is the solution of $\hbar \omega =\varepsilon _{+}(k_{\text{F}})-\varepsilon
_{-}(k_{\text{F}})$. We have set $k_{x}=k\cos \phi $, $k_{y}=\frac{t}{%
\lambda }k\sin \phi $ at the X point and $k_{x}=\frac{t}{\lambda }k\cos \phi 
$, $k_{y}=k\sin \phi $ at the Y point. We have also used $\partial
_{k}\varepsilon \left( k\right) =t^{2}a^{2}k/\sqrt{t^{2}a^{2}k^{2}+\Delta
^{2}}$.  The optical conductivity at the optical band edge $\hbar
\omega =2\left\vert \Delta \right\vert $  is given by%
\begin{equation}
\sigma \left( \omega =\frac{2\left\vert \Delta \right\vert }{\hbar }%
;\vartheta \right) =\frac{\hbar \sigma _{\text{H}}\left\vert \Delta
\right\vert }{a^{2}t^{2}}g(\mathbf{0};\vartheta ),  \label{OptD}
\end{equation}%
where we have used $k_{\text{F}}=0$ at the optical band edge.

The condition that no electrons are excited at the Y point is given by%
\begin{equation}
g^{\text{Y}}(\mathbf{0};\vartheta ^{\text{Y}})=0.  \label{gY}
\end{equation}%
By solving this equation, the ellipticity of the light $\vartheta ^{\text{Y}%
} $ is explicitly determined as%
\begin{equation}
\vartheta ^{\text{Y}}=\eta \arccos \frac{t}{\sqrt{\lambda ^{2}+t^{2}}},
\end{equation}%
with $\eta =\pm 1$, where $\eta =1$ for $\lambda \Delta >0$ and $\eta =-1$
for $\lambda \Delta <0$.  Then, it follows from Eq.(\ref{gk}) and Eq.(\ref%
{OptD}) that the optical conductivity at the optical band edge $\hbar
\omega =2\left\vert \Delta \right\vert $ is given by%
\begin{equation}
\frac{\sigma ^{\text{X}}\left( \omega =\frac{2\left\vert \Delta \right\vert 
}{\hbar };\vartheta ^{\text{Y}}\right) }{\frac{\hbar \sigma _{\text{H}%
}\left\vert \Delta \right\vert }{a^{2}t^{2}}}=\pi \frac{\left( \lambda
^{2}+\eta \frac{\lambda \Delta }{\left\vert \lambda \Delta \right\vert }%
t^{2}\right) ^{2}}{2t^{2}\left\vert \Delta \right\vert \left( \lambda
^{2}+t^{2}\right) },  \label{sM}
\end{equation}%
 implying that electrons are excited at the X point. Hence, elliptic
optical dichroism occurs.

In general, the $\omega $\ dependence of the optical conductivity is
obtained as%
\begin{equation}
\frac{\sigma ^{\text{Y}}\left( \omega ;\vartheta ^{\text{Y}}\right) }{\frac{%
\hbar \sigma _{\text{H}}\left\vert \Delta \right\vert }{a^{2}t^{2}}}=\frac{%
\pi \lambda ^{2}\left( \hbar \omega -\eta \frac{\lambda }{\left\vert \lambda
\right\vert }2\Delta \right) ^{2}}{\left( \hbar \omega \right) ^{3}\left(
\lambda ^{2}+t^{2}\right) }
\end{equation}%
 around the Y point, and%
\begin{align}
\frac{\sigma ^{\text{X}}\left( \omega ;\vartheta ^{\text{Y}}\right) }{\frac{%
\hbar \sigma _{\text{H}}\left\vert \Delta \right\vert }{a^{2}t^{2}}}=& \frac{%
\pi }{2\lambda t\left( \hbar \omega \right) ^{3}\left( \lambda
^{2}+t^{2}\right) }  \notag \\
& \hspace{-0.8in}\times \left( 8\eta \Delta \hbar \omega \left\vert \lambda
\right\vert \lambda t^{2}+\left( \hbar ^{2}\omega ^{2}+4\Delta ^{2}\right)
\left( \lambda ^{4}+t^{4}\right) \right)
\end{align}%
 around the X point. The optical conductivity is shown as a function of $%
\omega $ in Fig.\ref{FigOpt}(a). The optical conductivity takes the maximum
value (\ref{sM}) at the optical band edge $\hbar \omega =2\left\vert
\Delta \right\vert $  and monotonically decreases $\sigma ^{\text{X}%
}\left( \omega ;\vartheta ^{\text{Y}}\right) \propto 1/\omega $ as the
increase of $\omega $. 

Similarly, no electrons are excited at the X point if 
\begin{equation}
g^{\text{X}}(\mathbf{0};\vartheta ^{\text{X}})=0
\end{equation}%
with%
\begin{equation}
\vartheta ^{\text{X}}=\eta \arccos \frac{\lambda }{\sqrt{\lambda ^{2}+t^{2}}}%
,
\end{equation}
 where the optical conductivity at the optical band edge is given by%
\begin{equation}
\sigma \left( \omega =\frac{2\left\vert \Delta \right\vert }{\hbar }%
;\vartheta ^{\text{X}}\right) =\sigma _{0}g^{\text{Y}}(\mathbf{0};\vartheta
^{\text{X}})
\end{equation}%
with%
\begin{equation}
g^{\text{Y}}(\mathbf{0};\vartheta ^{\text{X}})=\pi \frac{\left( \lambda
^{2}+\eta \frac{\lambda \Delta }{\left\vert \lambda \Delta \right\vert }%
t^{2}\right) ^{2}}{2t^{2}\left\vert \Delta \right\vert \left( \lambda
^{2}+t^{2}\right) }.
\end{equation}%
 The optical conductivity is shown as a function of $\vartheta $ at the
optical band edge in Fig.\ref{FigOpt}(b).\emph{\ }

It is possible to make $g^{\text{Y}}(0;\vartheta ^{\text{Y}})=0$ or $g^{%
\text{X}}(0;\vartheta ^{\text{X}})=0$ by tuning the ellipticity of
irradiated light $\vartheta $, where only electrons at the X point or the Y
point are selectively excited. It is elliptic optical dichroism. 
\begin{figure}[t]
\centerline{\includegraphics[width=0.49\textwidth]{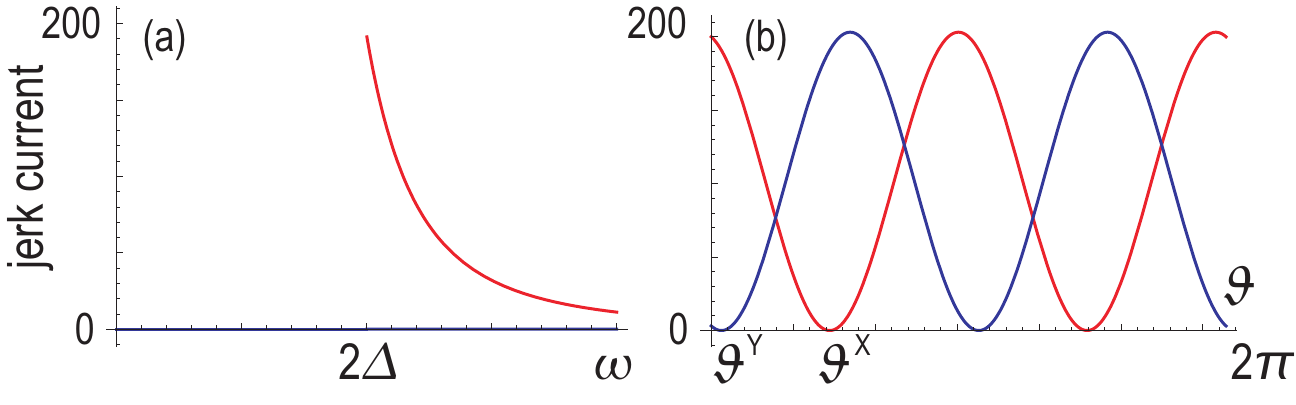}}
\caption{(a) Jerk current as a function of $\protect\omega $ dependence  %
in units of $\frac{2}{\left( i\protect\omega _{0}+1/\protect\tau \right) ^{2}%
}\frac{2\protect\pi e^{4}}{\hbar ^{3}V}E\left( \protect\omega \right)
E\left( -\protect\omega \right) E_{x}\left( 0\right) $.   Red (blue)
curve corresponds to the spin up (down) polarization. We have set $\protect%
\vartheta =\protect\vartheta ^{\text{Y}}$, where only electrons with up spin
is excited. (b) Jerk current as a function of $\protect\vartheta $ at the
optical band edge $\hbar \protect\omega =2\left\vert \Delta \right\vert $.}
\label{FigJerk}
\end{figure}

\textbf{Jerk current under elliptically polarized light:} We now show that
these excited electrons are extracted by applying electric field
additionally. The second-order photocurrents such as the injection current%
\cite{Sipe,JuanNC,Juan,Ave,AhnX,AhnNP,WatanabeInject,Okumura,Dai,EzawaVolta}
and the shift current\cite%
{Young,Young2,Ave,Kraut,Baltz,Sipe,Juan,AhnX,MorimotoScAd,Kim,Barik,AhnNP,WatanabeInject,Dai,Yoshida,EzawaVolta}
are zero due to inversion symmetry. On the other hand, the third-order
photocurrent\cite{Jerk,JerkComment,JerkReply,snap,PJerk} is not prohibited
by the presence of inversion symmetry. The higher-order injection current
was derived for the linearly polarized light\cite{PJerk}.  The injection
current $J_{\text{inject}}^{x;x2}$ originates in the velocity difference $%
v_{+}^{x}-v_{-}^{x}$, while $J_{\text{jerk}}^{x;x^{3}}$ originates in the
contribution $\frac{dv_{+}^{x}}{dt}-\frac{dv_{-}^{x}}{dt} $. In the similar
way, the $\ell $-th order photocurrent originates in 
\begin{equation}
\frac{d^{\ell -2}v_{+}^{x}}{dt^{\ell -2}}-\frac{d^{\ell -2}v_{-}^{x}}{%
dt^{\ell -2}}=\frac{\partial ^{\ell -1}\omega _{+-}}{\partial k_{x}^{\ell -1}%
}.
\end{equation}%
 We generalize it to elliptically polarized light, where the $\ell $-th
order photocurrent is \emph{\ }given by%
\begin{align}
J^{x;x^{\ell }}\left( \omega \right) =& \frac{\ell -1}{\left( i\omega
_{0}+1/\tau \right) ^{\ell -1}}\frac{2\pi e^{\ell +1}}{\hbar ^{\ell }V}\int d%
\mathbf{k}g\left( \mathbf{k};\vartheta \right) \delta \left( \omega
_{+-}-\omega \right)  \notag \\
& \times \frac{\partial ^{\ell -1}\omega _{+-}}{\partial k_{x}^{\ell -1}}%
E\left( \omega \right) E\left( -\omega \right) \left[ E_{x}\left( 0\right) %
\right] ^{\ell -2},  \label{jerk}
\end{align}%
where $\omega _{0}$ is the frequency of applied electric field and $\hbar
\omega _{+-}\equiv \left( \varepsilon _{+}-\varepsilon _{-}\right) $ with
Eq.(\ref{ek}) is the difference of the energy between the conduction and
valence bands. The detailed derivation is shown in Supplemental Material II.

We calculate the jerk current with $\ell =3$ at the optical band edge, which
is the leading photocurrent. It is obtained from Eq.(\ref{jerk}) as%
\begin{equation}
\frac{J^{x;x^{3}}\left( \Delta ;\vartheta ^{\text{Y}}\right) }{\frac{2}{%
\left( i\omega _{0}+1/\tau \right) ^{2}}\frac{2\pi e^{4}}{\hbar ^{3}V}%
E\left( \omega \right) E\left( -\omega \right) E_{x}\left( 0\right) }=\frac{%
\pi t\left( \lambda ^{2}-t^{2}\right) ^{2}}{2\Delta ^{2}\lambda \left(
\lambda ^{2}+t^{2}\right) }  \label{J3}
\end{equation}%
at the X point. This result is derived as follows. It follows from Eq.(\ref%
{ek}) that%
\begin{align}
\frac{\partial ^{2}\omega _{+-}\left( \mathbf{k}\right) }{\partial k_{x}^{2}}%
=& \frac{2a^{2}t^{2}}{\sqrt{\left( tak_{x}^{\prime }\right) ^{2}+\left(
\lambda ak_{y}\right) ^{2}+\Delta ^{2}}}  \notag \\
& -\frac{2t^{4}a^{4}k_{x}^{2}}{\left( \left( tak_{x}^{\prime }\right)
^{2}+\left( \lambda ak_{y}\right) ^{2}+\Delta ^{2}\right) ^{3/2}},
\end{align}%
 which leads to 
\begin{equation}
\frac{\partial ^{2}\omega _{+-}\left( \mathbf{0}\right) }{\partial k_{x}^{2}}%
=\frac{2a^{2}t^{2}}{\left\vert \Delta \right\vert }  \label{w0}
\end{equation}%
 at the optical band edge. In addition, it follows from Eqs.(\ref{g1}), (%
\ref{g2}) and (\ref{omega}) that%
\begin{equation}
g_{xx}^{\text{X}}\left( \mathbf{0}\right) =\frac{a^{2}t^{2}}{4\Delta ^{2}}%
,\quad g_{yy}^{\text{X}}\left( \mathbf{0}\right) =\frac{a^{2}\lambda ^{2}}{%
4\Delta ^{2}},\quad \Omega _{xy}^{\text{X}}\left( \mathbf{0}\right) =\frac{%
a^{2}t\lambda \Delta }{2\Delta ^{3}}.
\end{equation}%
 With the use of these and Eq.(\ref{gk}), we obtain 
\begin{equation}
g\left( \mathbf{0};\vartheta \right) =\frac{a^{2}t^{2}}{4\Delta ^{2}}\cos
^{2}\vartheta +\frac{a^{2}\lambda ^{2}}{4\Delta ^{2}}\sin ^{2}\vartheta +%
\frac{a^{2}t\lambda \Delta }{2\Delta ^{3}}\sin \vartheta \cos \vartheta .
\label{g0}
\end{equation}%
 By inserting Eq.(\ref{w0}) and Eq.(\ref{g0}) to Eq.(\ref{jerk}), we
obtain Eq.(\ref{J3}). It is interesting that the jerk current is zero for
isotropic Dirac cone with $\lambda =t$ at the optical band edge as in Eq.(%
\ref{J3}).

In general, the $\omega $\ dependence of the jerk current is explicitly
given by%
\begin{align}
& \frac{J^{x;x^{3}}\left( \omega ;\vartheta ^{\text{Y}}\right) }{\frac{2}{%
\left( i\omega _{0}+1/\tau \right) ^{2}}\frac{2\pi e^{4}}{\hbar ^{3}V}%
E\left( \omega \right) E\left( -\omega \right) E_{x}\left( 0\right) }  \notag
\\
=& \frac{a^{2}\pi t}{2\left( \hbar \omega \right) ^{6}\left( \lambda
^{2}+t^{2}\right) }  \notag \\
& \times \Big[\left( \lambda ^{4}+3t^{4}\right) \left( \left( \hbar \omega
\right) ^{4}+16\Delta ^{4}\right) +8\Delta ^{2}\left( 3\lambda
^{4}+t^{4}\right) \left( \hbar \omega \right) ^{2}  \notag \\
& +16\hbar \omega \Delta \lambda ^{2}t^{2}\left( \left( \hbar \omega \right)
^{2}+4\Delta ^{2}\right) \Big]
\end{align}%
 at the X point, and%
\begin{align}
& \frac{J^{x;x^{3}}\left( \omega ;\vartheta ^{\text{Y}}\right) }{\frac{2}{%
\left( i\omega _{0}+1/\tau \right) ^{2}}\frac{2\pi e^{4}}{\hbar ^{3}V}%
E\left( \omega \right) E\left( -\omega \right) E_{x}\left( 0\right) }  \notag
\\
=& \frac{2\pi \left( a\lambda t\right) ^{2}\left( \hbar ^{2}\omega
^{2}+4\Delta ^{2}\right) \left( \hbar \omega +2\Delta \right) ^{2}}{\left(
\hbar \omega \right) ^{6}\left( \lambda ^{2}+t^{2}\right) }
\end{align}%
 at the Y point.

The jerk current is shown as a function of $\omega $ in Fig.\ref{FigJerk}%
(a), where the jerk current takes the maximum value (\ref{J3}) at the
optical band edge $\hbar \omega =2\left\vert \Delta \right\vert $ and
monotonically decreases $J_{\text{inject}}^{x;x^{3}}\left( \omega ;\vartheta
^{\text{Y}}\right) \propto 1/\omega ^{2}$ as the increase of $\omega $. The
jerk current with up spins is induced by elliptically polarized light with $%
\vartheta =\vartheta ^{\text{Y}}$, while there is no jerk current with down
spin at the optical band edge $\hbar \omega =2\left\vert \Delta \right\vert $%
. We show the jerk current as a function of $\vartheta $ at the optical band
edge in Fig.\ref{FigJerk}(b). The portion of the spin polarization is
controlled by tuning the ellipticity $\vartheta $.

We note that, in the case of the isotropic Dirac cone with $\lambda =t$, the
jerk current is identical at the X and Y points, and given by%
\begin{align}
& \frac{J^{x;x^{3}}\left( \omega ;\vartheta ^{\text{Y}}\right) }{\frac{2}{%
\left( i\omega _{0}+1/\tau \right) ^{2}}\frac{2\pi e^{4}}{\hbar ^{3}V}%
E\left( \omega \right) E\left( -\omega \right) E_{x}\left( 0\right) }  \notag
\\
=& \frac{a^{2}\pi t^{2}\left( \hbar ^{2}\omega ^{2}+4\Delta ^{2}\right)
\left( \hbar \omega +2\Delta \right) ^{2}}{\left( \hbar \omega \right) ^{6}}.
\end{align}%
 The jerk current is zero at the optical band edge for the isotropic
Dirac cones,%
\begin{equation}
J^{x;x^{3}}\left( \omega =\frac{2\left\vert \Delta \right\vert }{\hbar }%
;\vartheta ^{\text{Y}}\right) =0.
\end{equation}%
 Hence, the jerk current at the optical band edge is present only for
anisotropic Dirac cones.

\textbf{Discussions:} We have shown that the low-energy physics of $d$-wave
altermagnets is governed by anisotropic Dirac cones at the X and Y points,
where the electronic spins are perfectly polarized in opposite directions.
This unique band structure enables complete spin selectivity under
elliptically polarized light, giving rise to elliptic optical dichroism.
When static electric field is applied simultaneously with elliptically
polarized light, a perfectly spin-polarized jerk current is generated.
Notably, this jerk current exhibits a pronounced enhancement at the optical
band edge, a feature that appears only in systems with anisotropic Dirac
cones.  We have demonstrated that insulating $d$-wave altermagnets host a
spin-polarized third-order photocurrent as their leading electric current.
These findings establish $d$-wave altermagnets as a promising platform for
photoinduced spintronics.

The perfect spin selectivity identified here originates in the spin-split
band structure with fully diagonal spin eigenstates. This stands in sharp
contrast to Dirac cones arising from spin--orbit interactions such as those
in Rashba systems or topological-insulator surfaces, where the spin texture
is non-diagonal and perfect selectivity is impossible. Altermagnets, by
realizing large spin-split bands without spin-orbit coupling, provide an
ideal setting in which opposite spin polarizations at the two valleys are
protected by $\mathcal{C}_{4z}\mathcal{T}$ symmetry. By the same reasoning,
we expect analogous spin-selective optical phenomena in $g$-wave and $i$%
-wave altermagnets owing to $\mathcal{C}_{8z}\mathcal{T}$ \ and $\mathcal{C}%
_{12z}\mathcal{T}$ symmetries, respectively, as well.

Spin-polarized current is experimentally observable by means of the
tunneling magneto-resistance (TMR)\cite{Julli,Moodera}. Especially,
perfectly spin-polarized current was observed by using the TMR\cite%
{Park,Bowen,Yuasa}. In addition, the spin-polarized scanning tunneling
spectroscopy\cite{Machida} and point-contact Andreev reflection\cite{Lee}
are used for the detection of spin-polarized current. 

This work is supported by CREST, JST (Grants No. JPMJCR20T2) and
Grants-in-Aid for Scientific Research from MEXT KAKENHI (Grant No. 23H00171).

\end{document}